\documentclass[preprint,a4paper]{elsarticle}

\pdfoutput=1

\usepackage{lineno}
\usepackage[utf8]{inputenc}
\usepackage[verbose]{placeins}
\usepackage{amsmath}
\usepackage{amssymb}
\usepackage{amsfonts,mathrsfs,fixmath,amsthm}
\usepackage{bbm}
\usepackage{algpseudocode}
\usepackage{algorithm}

\newcommand{\Dov}{\ensuremath{D_\text{ov}}}
\newcommand{\Dw}{\ensuremath{D_\text{w}}}

\newcommand{\ket}[1]{\ensuremath{\left|#1\right\rangle}}
\newcommand{\bra}[1]{\ensuremath{\left\langle#1\right|}}
\newcommand{\braket}[2]{\ensuremath{\left\langle\left.#1\right| #2
    \right\rangle}}

\newcommand{\RR}[1]{\ensuremath{\ket{R_{#1}}}}
\newcommand{\LL}[1]{\ensuremath{\bra{L_{#1}}}}
\newcommand{\dRR}[1]{\ensuremath{\ket{\pt R_{#1}}}}
\newcommand{\dLL}[1]{\ensuremath{\bra{\pt L_{#1}}}}
\newcommand{\pt}{\ensuremath{\partial_{t}}}

\newcommand{\bA}{\ensuremath{\mathcal{B}}}
\newcommand{\upperRomannumeral}[1]{\uppercase\expandafter{\romannumeral#1}}
\newcommand{\Tr}{\ensuremath{\operatorname{tr}}}
\newcommand{\CC}{\ensuremath{\mathbb{C}}}
\newcommand{\CNN}{\ensuremath{\mathbb{C}^{n \times n}}}
\newcommand{\id}{\ensuremath{\mathbbm{1}}}
\newcommand{\MeV}{\ensuremath{\text{MeV}}}

\DeclareMathOperator{\sgn}{sgn}
\DeclareMathOperator{\TSL}{TSL}
\DeclareMathOperator{\diag}{diag}
\renewcommand{\Re}{\ensuremath{\operatorname{Re}}}

\newcommand{\lr}[1]{\left(#1\right)}
\DeclareMathOperator{\fm}{fm}

\newtheorem{theorem}{Theorem}

\journal{Elsevier}

\graphicspath{ {./plots_dir/} {./plots/} {./} }

\biboptions{numbers,sort&compress}

\begin{document}
\sloppy

\begin{frontmatter}

\title{A numerical method to compute derivatives of functions of large complex matrices
  and its application to the overlap Dirac operator at finite chemical potential}



\author[address]{Matthias Puhr}
\ead{Matthias.Puhr@physik.uni-regensburg.de}

\author[address]{Pavel Buividovich} 
\ead{Pavel.Buividovich@physik.uni-regensburg.de}

\address[address]{Institute of Theoretical Physics, Regensburg University, D-93053 Regensburg, Germany}

\begin{abstract}
 We present a method for the numerical calculation of derivatives of functions of general
complex matrices. The method can be used in combination with any algorithm that evaluates
or approximates the desired matrix function, in particular with implicit Krylov--Ritz-type
approximations. An important use case for the method is the evaluation of the overlap
Dirac operator in lattice Quantum Chromodynamics (QCD) at finite chemical potential, which
requires the application of the sign function of a non-Hermitian matrix to some source
vector. While the sign function of non-Hermitian matrices in practice cannot be
efficiently approximated with source-independent polynomials or rational functions,
sufficiently good approximating polynomials can still be constructed for each particular
source vector. Our method allows for an efficient calculation of the derivatives of such
implicit approximations with respect to the gauge field or other external parameters,
which is necessary for the calculation of conserved lattice currents or the fermionic
force in Hybrid Monte-Carlo or Langevin simulations. We also give an explicit deflation
prescription for the case when one knows several eigenvalues and eigenvectors of the
matrix being the argument of the differentiated function. We test the method for the
two-sided Lanczos approximation of the finite-density overlap Dirac operator on realistic
$SU(3)$ gauge field configurations on lattices with sizes as large as $14\times14^3$ and
$6\times18^3$.
\end{abstract}


\end{frontmatter}

\section{Introduction}
\label{sec:Intro}

 Over the last decade matrix valued functions of matrices have become an essential tool in
a variety of sub-fields of science and engineering \cite{Higham2008}.  An important
application for matrix functions in the field of lattice QCD is the so-called overlap
Dirac operator, which is a discretisation of the Dirac operator that respects the properly
defined lattice chiral symmetry (Ginsparg-Wilson relations) and is free of
doublers. Therefore the overlap Dirac operator is well suited for the non-perturbative
study of strongly interacting chiral fermions. At finite chemical potential $\mu$ the
overlap Dirac operator is defined as \cite{Bloch:06:1}
\begin{equation}
\label{eq:overlap}
 \Dov := \frac{1}{a}\left(\mathbbm{1}+\gamma_5 \sgn\left[ H(\mu)
\right] \right),
\end{equation} where $H(\mu) := \gamma_5 \Dw(\mu)$, $\Dw(\mu)$ is the Wilson--Dirac
operator at chemical potential $\mu$, $\sgn$ is the matrix sign function and $a$ stands
for the lattice spacing. An explicit form of $\Dw(\mu)$ is given in \ref{ap:WD_operator}.

 At finite chemical potential $H$ is a non-Hermitian matrix with complex
eigenvalues. Since the size of the linear space on which $H$ is defined is typically very
large ($n \sim 10^4 \ldots 10^7$), it is not feasible to evaluate the matrix sign function
exactly. While at $\mu = 0$ one can efficiently approximate the sign function of the
Hermitian operator $H$ by polynomials or rational functions
\cite{Cundy:11:1,Giusti2002,van_den_Eshof:02:01,Kennedy:04:01}, at nonzero $\mu$ the
operator $H$ becomes non-Hermitian and such approximations typically become
inefficient. However, having in mind that in practice the sign function
$\sgn\left[H(\mu)\right]$ is applied to some source vector, one can still construct an
efficient polynomial approximation for each particular source vector by using Krylov
subspace methods, such as Krylov--Ritz-type approximations. One of the practical
Krylov--Ritz-type approximations which are suitable for the finite-density overlap Dirac
operator is the two-sided Lanczos (TSL) algorithm, developed in
\cite{Bloch2010,Bloch2008}. The efficiency of the TSL approximation can be further
improved by using a nested version of the algorithm \cite{Bloch:11:2}.

 Many practical tasks within lattice QCD simulations require the calculation of the
derivatives of the lattice Dirac operator with respect to the gauge fields or some other
external parameters. For example, conserved lattice vector currents and fermionic force
terms in Hybrid Monte-Carlo simulations involve the derivatives of the Dirac operator with
respect to Abelian or non-Abelian gauge fields. Also, electric charge susceptibilities
which are used to quantify electric charge fluctuations in quark--gluon plasma involve
derivatives of the Dirac operator with respect to the chemical potential.

 While explicit expressions for the derivatives of source-independent approximations of
matrix functions are well known and are routinely used in practical lattice QCD
simulations, differentiating the implicit source-dependent approximation appears to be a
more subtle problem. In principle the algorithms for taking numerical derivatives of
scalar functions, like the finite difference method or algorithmic differentiation, can be
generalised to matrix functions and to matrix function approximation algorithms. It is
easy to combine an approximation with the finite difference method, but finite differences
are very sensitive to round-off errors and it is often not possible to reach the desired
precision in the derivative using this method. For algorithmic differentiation the
situation is more complicated. Depending on the approximation method used it might not be
immediately clear how to apply algorithmic differentiation. Even if algorithmic
differentiation can be implemented for the approximation method this might lead to a
numerically unstable algorithm. Such a behaviour was observed when the TSL approximation
was used in conjunction with algorithmic differentiation~\cite{Buividovich:14:3}.

 In this paper, we propose and test a practical numerically stable method which makes it
possible to compute derivatives of implicit approximations of matrix functions to high
precision. The main motivation for this work is the need to take derivatives of the
overlap Dirac operator in order to compute conserved currents on the lattice.

 The structure of the paper is the following: in Section \ref{sec:mat_func} we state some
general theorems about matrix functions and their derivatives, which provide the basis for
our numerical method. In Section \ref{sec:deflation} we discuss how the calculation of the
derivatives of matrix functions can be made more efficient by deflating a number of small
eigenvalues of the matrix which is the argument of the function being differentiated. The
deflation is designed with the matrix sign function and the TSL approximation in
mind. Nevertheless we want to emphasise that the method is very general and can be applied
to other matrix functions and different matrix function approximation schemes. In Section
\ref{sec:results} we demonstrate how the method can be used in practice. First we discuss
the efficiency and convergence of the TSL approximation. After that, as a test case, we
compute $U\lr{1}$ lattice vector currents, which involve the derivatives of the overlap
Dirac operator with respect to background Abelian gauge fields, and demonstrate that they
are conserved. Finally we summarise and discuss the advantages and disadvantages of our
method in Section \ref{sec:discussion}. Detailed calculations and derivations as well as a
pseudo code implementation of the method can be found in the Appendices.

\section{Matrix functions and numerical evaluation of their derivatives}
\label{sec:mat_func}

 For completeness we start this Section with a brief review of matrix functions. Let the
function \mbox{$f: \CC \to \CC$} be defined on the spectrum of a matrix $A \in
\CNN$. There exist several equivalent ways to define the generalisation of $f$ to a matrix
function $f: \CNN \to \CNN$\cite{Higham2008,Golub1996}. For the purpose of this paper the most
useful definition is via the Jordan canonical form. A well known
Theorem states that any matrix $A \in \CNN$ can be written in the Jordan
canonical form
\begin{equation}
  \label{eq:A_Jordan}
  X^{-1} A X = J = \diag\left(J_1,J_2,\dots,J_k\right),
\end{equation}
where every Jordan block $J_i$ corresponds to an eigenvalue $\lambda_i$ of $A$ and has the
form
\begin{equation}
  \label{eq:Jordan_block}
  J_i = J_i(\lambda_i) = \begin{pmatrix}
    \lambda_i & 1         &   0     &  \cdots    & 0       \\
         0    & \lambda_i &   1     &  \ddots    &  \vdots \\
         0    & \ddots    & \ddots  &  \ddots    &  0      \\
      \vdots  & \ddots    & \ddots  &  \lambda_i &  1      \\
          0   & \cdots    & 0       &    0       & \lambda_i
  \end{pmatrix} \in \CC^{m_i \times m_i},
\end{equation}
with $m_1 + m_2 + \dots + m_k = n$. The Jordan matrix $J$ is unique up to permutations of
the blocks but the transformation matrix $X$ is not.  Using the Jordan canonical form the
matrix function can be defined as \cite{Higham2008,Golub1996}
\begin{equation}
  \label{eq:matrix_func}
   f(A) := X f(J) X^{-1} = X \diag(f(J_i)) X^{-1}.
\end{equation}
The function of the Jordan blocks is given by
\begin{equation}
  \label{eq:Jordan_function}
  f(J_i) := \begin{pmatrix}
    f(\lambda_i) &  f'(\lambda_i)  &  \dots   & \frac{f^{(m_i-1)}(\lambda_i)}{(m_i-1)!}  \\
          0      & f(\lambda_i)    &  \ddots  & \vdots                                  \\
    \vdots       &    \ddots       &  \ddots  & f'(\lambda_i)                           \\
          0      &    \cdots       &  0        & f(\lambda_i)
  \end{pmatrix} \in \CC^{m_i \times m_i},
\end{equation}
Note that this definition requires the existence of the derivatives
$f^{(m_i-1)}(\lambda_i)$ for $i = 1, \dots , k$. If $A$ is diagonalisable every Jordan
block has size one and equation~(\ref{eq:Jordan_function}) reduces to the so-called
spectral form
\begin{equation}
  \label{eq:spectral_form}
  f(A) = X \diag(f(\lambda_1), f(\lambda_2), \dots, f(\lambda_n)) X^{-1} ,
\end{equation}
which does not depend on the derivatives of $f$. For instance, the matrix sign function $\sgn\left[ H(\mu)
\right]$ in (\ref{eq:overlap}) is defined by $\sgn(\lambda_i) = \sgn\lr{\Re \lambda_i}$.

 Finding the Jordan normal form (or the spectral decomposition) of a matrix is a
computationally expensive task and takes $\mathcal{O}(n^3)$ operations. For large matrices
it is therefore not feasible to compute the matrix function exactly. In practical
calculations it is often sufficient to know the result $\ket{y}=f(A)\ket{x}$ of
the action of the matrix function on a vector and it is not necessary to explicitly
compute the matrix $f(A)$. A variety of different methods have been
developed to efficiently calculate an approximation of $\ket{y}=f(A)\ket{x}$
(see chapter 13 of \cite{Higham2008} for an overview). For non-Hermitian matrices $A$ one
typically constructs the approximation to $\ket{y}$ using the Krylov subspaces spanned by the
Krylov vectors $A^{k} \ket{x}$ and $\lr{A^{\dag}}^k \ket{x}$, $k = 0, \ldots,
k_{max}$. This implies that the approximation depends both on the matrix $A$ itself and on
the source vector $\ket{x}$. A practical example of such an approximation is the TSL
algorithm of \cite{Bloch2010,Bloch2008,Bloch:11:2}.

Let us now assume that the matrix $A \equiv A\lr{t}$ depends on some
parameter $t$. In this work we are interested in the calculation of the derivative
\begin{equation}
\label{eq:derivative_def}
 \partial_t \ket{y} = \lr{\partial_t f\lr{A\lr{t}}} \ket{x} ,
\end{equation}
where $\partial_t \equiv \frac{\partial}{\partial t}$ and the source vector
$\ket{x}$ is assumed to be independent of $t$. The practical method for calculating
$\partial_t \ket{y}$ which we propose here is based on the following Theorem \cite{Mathias:96:1}:
\begin{theorem}
\label{th:mathias}
   Let $A(t) \in \mathbb{C}^{n\times n}$ be differentiable at $t=0$ and assume that the
spectrum of $A(t)$ is contained in an open subset $\mathcal{D}\subset\mathbb{C}$ for all
$t$ in some neighbourhood of $0$. Let $f$ be $2n-1$ times continuously differentiable
on~$\mathcal{D}$. Then
    $$ f\left(\bA\right) \equiv \left(\begin{array}{cc}
        f(A(0)) & \left. \pt  f(A(t)) \right|_{t=0} \\
        0 & f(A(0))
      \end{array}\right), \quad \bA(A) := \left(
        \begin{array}{cc}
          A(0) & \left. \pt A(t) \right|_{t=0} \\
          0   & A(0)
        \end{array}
      \right) $$
\end{theorem}
Theorem \ref{th:mathias} relates the derivative of a matrix function $\pt f(A)$ to the function of a
block matrix $f(\bA(A))$, and is in fact closely related to the formula (\ref{eq:Jordan_function}). It
is remarkable that one does not need to know the explicit form of $\pt f$ to
compute the derivative of $f$. This comes at the cost of evaluating the function $f$ for
the matrix $\bA$ that has twice the dimension of $A$ and one needs to know the derivative
$\pt A$. In practice $\pt A$ is usually known analytically or can be computed to high
precision. Moreover the block matrix $\bA$ is very sparse and one only needs to store $A$
and $\pt A$. Thus Theorem \ref{th:mathias} makes it possible to efficiently calculate
the derivative of a matrix function.

 Using Theorem \ref{th:mathias} it is straightforward to compute the action of the derivative of a
matrix function on a vector:
\begin{equation}
  \label{eq:func_derivative_1}
  f(\bA) \begin{pmatrix} 0 \\ \ket{x} \end{pmatrix} = \begin{pmatrix}
    \pt f(A) \ket{x} \\ \ \ \  f(A) \ket{x}  \end{pmatrix}
\end{equation}
Now one can use any matrix function approximation method to compute an approximation to
the function $f(\bA)$ in equation (\ref{eq:func_derivative_1}). To be specific we will use
the TSL approximation for the rest of this paper, but any other approximation scheme can
be used instead.

\section{Spectral properties of the block matrix $\bA$ and deflation of the derivatives of
matrix functions}
\label{sec:deflation}

 The convergence properties of the TSL approximation crucially depend on the spectrum of
the matrix to which it is applied. In \ref{ap:prop_mathias} we show that the spectrum of
eigenvalues of $\bA$ is identical to the spectrum of $A$.  Often the efficiency of the TSL
approximation can be greatly improved by deflating a small number of eigenvalues. One
might for example deflate the eigenvalues that are close to a pole of the function. For
the matrix sign function it is advantageous to deflate the eigenvalues smallest in
absolute value\cite{Bloch2008,Bloch2010, Bloch:11:2}. The standard deflation method relies
on the diagonalisability of the matrix $A$. It is then straightforward to project out the
eigenvectors corresponding to the deflated eigenvalues.

However the following Theorem states that, in general, the  matrix $\bA$ is not
diagonalisable:
\begin{theorem}
  \label{th:diag} Let $A$ be a diagonalisable matrix. If $\pt \lambda_i \neq 0 $ for at
least one eigenvalue $\lambda_i$ of
$A$ then the matrix $\bA$ is not diagonalisable. If $A$ has no degenerate eigenvalues and
$\pt \lambda_i \neq 0 $ for all $i \in \{1, \dots, n\}$ then every Jordan block in the Jordan
normal form of $\bA$ is of size two, i.e.
  $$ \mathcal{J}  = \begin{pmatrix}
    J_1 & 0 & 0 & \dots & 0 \\ 0 & J_2 & \ddots & \ddots & \vdots \\ 0 & \ddots & \ddots &
\ddots & 0 \\ \vdots & \ddots & \ddots & \ddots & 0 \\ 0 & \dots & 0 & 0 & J_n
  \end{pmatrix}, \quad J_i:=
  \begin{pmatrix} \lambda_i & 1 \\ 0 & \lambda_i
  \end{pmatrix}.
$$
\end{theorem}
\noindent The proof of this Theorem is somewhat technical and the main steps of the proof can be found in
\mbox{\ref{ap:prop_mathias}}. It follows directly from Theorem \ref{th:diag} that
a matrix of the form $\bA$ does not possess a full basis of eigenvectors and it is therefore not
immediately clear how to apply deflation to the numerical evaluation of
equation~(\ref{eq:func_derivative_1}).

In the following we develop a deflation method that is based on the Jordan normal form of
$\bA$. Since $\mathcal{J}$ is the Jordan matrix of $\bA$ there exists
an invertible matrix $\mathcal{X}$ such that $\mathcal{X}^{-1} \bA \mathcal{X} = \mathcal{J}$. Using Theorem
\ref{th:diag} it is possible to derive an analytic expression for $\mathcal{X}$ and $\mathcal{X}^{-1}$ in
terms of the eigenvalues and left and right eigenvectors of $A$ and their
derivatives. After some algebraic manipulations that are summarised in
\ref{ap:jordan_decomp}  we get

\begin{eqnarray}
\label{eq:xinv_final}
  \mathcal{X}^{-1} & = &
  \begin{pmatrix}
    \left(\LL{1}, {\dLL{1}} \right) \\
    {\pt \lambda}_1 \left(0, \LL{1}\right) \\
     \vdots  \\
     \left(\LL{n}, {\dLL{n}} \right)  \\
     {\pt \lambda}_n \left(0, \LL{n}\right)
  \end{pmatrix} \qquad \text{and} \\
  \nonumber &  & \\
\label{eq:x_final}
  \mathcal{X} & = &
  \begin{pmatrix}
    \begin{pmatrix}
      \RR{1} \\
       0
    \end{pmatrix} \! \! \! \! &
    ,\frac{1}{{\pt \lambda}_1}
    \begin{pmatrix}
      {\dRR{1}} \\
      \RR{1}
    \end{pmatrix}  \! \! \! \!  &\! \! \!,\cdots,\! \! \! &
    \begin{pmatrix}
      \RR{n} \\
       0
    \end{pmatrix}   \! \! \! \! &
    ,  \frac{1}{{\pt \lambda}_n}
    \begin{pmatrix}
      {\dRR{n}} \\
      \RR{n}
    \end{pmatrix} \!
  \end{pmatrix},
\end{eqnarray}
where we used the eigenvalues $\lambda_i$ and  left (right) eigenvectors $\LL{i}$
$(\RR{i})$ of $A$. The
matrix $\mathcal{X}^{-1}$ consists of $2n$-dimensional row vectors, i.e.
\begin{equation}
\left(\LL{i}, {\dLL{i}}
\right)  = (\LL{i}_1, \dots, \LL{i}_n,\dLL{i}_1, \dots, \dLL{i}_n)\label{eq:Xinv_rows}.
\end{equation}
Analogously $\mathcal{X}$ is made up by $2n$-dimensional column vectors. With the help of the
matrices $\mathcal{X}$ and $\mathcal{X}^{-1}$ we can define a generalised deflation. Let $\ket{x_i}$ be
the columns of $\mathcal{X}$ and $\bra{\bar{x}_i}$ the rows of $\mathcal{X}^{-1}$, then
$\sum\limits_{i=1}^{2n} \ket{x_i}\bra{\bar{x}_i}=\mathbbm{1}$.  We define the projectors
$\mathcal{P}_m = \sum\limits_{i=1}^{m} \ket{x_i}\bra{\bar{x}_i}$  and
$\bar{\mathcal{P}}_m :=\mathbbm{1} - \sum\limits_{i=1}^{m} \ket{x_i}\bra{\bar{x}_i}$, such that for every
vector $\ket{\psi}$ we have
\begin{equation}
\label{eq:projector_1} \ket{\psi} = \mathcal{P}_m \ket{\psi}+\bar{\mathcal{P}}_m \ket{\psi}
\end{equation}
Now we have all the necessary ingredients to define a deflation algorithm for
the matrix $\bA$. Say we want to deflate the first $2l$ eigenvalues of $\bA$ (corresponding to
the first $l$ eigenvalues $\lambda_1, \dots, \lambda_l$ of $A$). To calculate
$f(\bA)$ we split the evaluation of the function into two parts:
\begin{equation}
 f(\bA)\ket{\psi} = \mathcal{X}f(\mathcal{J}) \mathcal{X}^{-1} \mathcal{P}_{2l} \ket{\psi}
+f(\bA)\bar{\mathcal{P}}_{2l} \ket{\psi} \hskip
 0.13\textwidth\label{eq:defl_1}
\end{equation}
where we used equation (\ref{eq:matrix_func}) for the first term on the right side. To
compute the function of the Jordan matrix we only need to consider the function for each
Jordan block. Applying equation (\ref{eq:Jordan_function}) to the $j$-th block of
$\mathcal{J}$ yields:
\begin{equation}
  \label{eq:defl_block}
  f(J_i) := \begin{pmatrix}
    f(\lambda_j) &  \partial_t f(\lambda_j)    \\
           0     & f(\lambda_j)
  \end{pmatrix}
\end{equation}
Since the block structure of $\mathcal{J}$ is preserved by the function and the
$\ket{x_i}$ and $\bra{\bar{x}_i}$ are biorthogonal one finds:
\begin{multline}
  \label{eq:defl_res_1} \mathcal{X} f(\mathcal{J}) \mathcal{X}^{-1} \mathcal{P}_{2l} \ket{\psi} = \\
\sum\limits_{i=1}^{l}\left[f(\lambda_i)\left(\ket{x_{2i-1}}\braket{\bar{x}_{2i-1}}{\psi} +
\ket{x_{2i}}\braket{\bar{x}_{2i}}{\psi}\right)+ (\pt f(\lambda_i))
\ket{x_{2j-1}}\braket{\bar{x}_{2i}}{\psi}\right]
\end{multline} It is not necessary to compute the full transformation matrices
$\mathcal{X}$ and $\mathcal{X}^{-1}$ since the rows and columns needed to evaluate
equation (\ref{eq:defl_res_1}) are known analytically in terms of the left and right
eigenvectors of $A$. The latter can be efficiently computed using the Arnoldi algorithm
(we have used the {\tt{ARPACK}} implementation \cite{Lehoucq1998}). In practical
calculations one chooses $2l \ll 2n$ and splits the calculation of $f(\bA)$ in the
following way
\begin{equation}
  \label{eq:final_res}
  \begin{split}
    f(\bA)\ket{\psi} & = \underbrace{f(\bA)\bar{\mathcal{P}}_{2l}\ket{\psi}}_{\text{TSL
        approximation}} + \\
                     &  \underbrace{\sum\limits_{i=1}^{l}\left[f(\lambda_i)\left(\ket{x_{2i-1}}\braket{\bar{x}_{2i-1}}{\psi} +
\ket{x_{2i}}\braket{\bar{x}_{2i}}{\psi}\right)+ (\pt f(\lambda_i))
\ket{x_{2i-1}}\braket{\bar{x}_{2i}}{\psi}\right]}_{\text{exact}}
  \end{split}
\end{equation}
This looks very similar to the standard deflation formula for diagonalisable matrices. The
difference is that now, because the Jordan blocks have size two, there is an additional
``mixing term'' proportional to $\partial_t f(\lambda_i)$. For the sign function equation~(\ref{eq:final_res})
becomes even simpler.  The sign function is piece-wise constant and for realistic gauge
field configurations the eigenvalues of $H$ do not cross the discontinuity line $\Re(z) =
0$ when an external parameter is varied. Therefore the derivative of the sign function
$\partial_t \sgn(\lambda_i)$ is identically zero and the mixing term is absent in the
deflation. A detailed explicit expression for the deflated derivative of the matrix sign
function and the corresponding pseudo-code can be found in \ref{ap:efficient_defl}.

We note that in Hybrid Monte-Carlo simulations with dynamical overlap fermions it is
possible that an eigenvalue $\lambda_i$ of $H$ crosses the discontinuity of the sign
function at $\Re{\lambda_i}=0$, which corresponds to a change of the topological charge.
The derivative of $\sgn\lr{\lambda_i}$ in the last term in (\ref{eq:final_res}) then
becomes singular.  However, in practice such singularities in the fermionic force are
typically avoided by modifying the Molecular Dynamics process in the vicinity of the
singularity, for example by using the transmission-reflection step of
\cite{Cundy:09:1,Fodor:04:01}.

\section{Numerical Tests for the overlap Dirac Operator}
\label{sec:results}

\subsection{Numerical setup and tuning of the TSL algorithm}
\label{subsec:setup_tuning}

 For the numerical tests we have used quenched $SU(3)$ gauge configurations generated with
the tadpole-improved Lüscher--Weisz action~\cite{Luescher1985,Gattringer:01:1}. The
Wilson--Dirac operator with a background Abelian gauge field and finite quark chemical
potential $\mu$ is described in \ref{ap:WD_operator}. There we also give an analytic
expression for the derivative $\frac{\partial \Dw(\mu)}{\partial \Theta_{x,\mu}}$ of the
Wilson--Dirac operator with respect to the external lattice gauge field $\Theta_{x,\mu}$.

 Before we discuss our results we give a detailed description of our numerical setup. The
results in this Section were obtained using a nested version of the TSL
algorithm~\cite{Bloch:11:2}. We found that the main performance gains are already achieved
with a single nesting step and that further nesting does not significantly improve the
efficiency of the algorithm. Therefore for all our calculations we used only one level of
nesting. In this case one has to choose two parameters for the Lanczos approximation, the
sizes of the inner and outer Krylov subspace. For the TSL method it is not known how to
compute a priori error estimates. In particular it is not possible to estimate which
values for the Krylov subspace size parameters are necessary to reach a given precision in
the calculation.  An a~posteriori estimate for the numerical error $\epsilon$ can be
computed using the
identity $\sgn(A)^2= \mathbbm{1}$:
\begin{equation}
  \label{eq:error_est}
   \epsilon_A = \frac{\| \sgn(A)^2 \ket{\psi} - \ket{\psi} \|}{2 \ \| \ket{\psi} \|},
\end{equation}
where $\|\ket{\psi}\| := \sqrt{\braket{\psi}{\psi}}$  and the factor two was added because
we have to apply the TSL approximation twice to compute the square of the sign function.

Similarly, in order to estimate the error $\epsilon_{\bA}$ in the calculation of the 
derivative of the sign function we use the same formula as (\ref{eq:error_est}) where $A$ is
replaced by $\bA$ and the vector $\ket{\psi}$ is replaced by the sparse vector $\left(0,
\ket{\psi}\right)^T$, i.e 

\begin{equation}
  \label{eq:bA_error_est}
   \epsilon_\bA = \frac{\left\| \sgn(\bA)^2
     \begin{pmatrix}
       0 \\
       \ket{\psi}
     \end{pmatrix}
 - \begin{pmatrix}
       0 \\
       \ket{\psi}
     \end{pmatrix} \right\|}{2 \ \left\| \ket{\psi} \right\|}.
\end{equation}
We note that the square of the sign of the block matrix $\bA$ is given by
\begin{equation}
  \label{eq:bA_squared}
  \sgn(\bA)^2 =
  \begin{pmatrix}
    \sgn(A)^2 & \sgn(A)(\pt \sgn(A)) + (\pt \sgn( A))\sgn(A) \\
        0  & \sgn(A)^2
  \end{pmatrix}
\end{equation}
and the error estimate for the derivative as defined in equation (\ref{eq:error_est}) contains  the
anti-commutator $\{\sgn(A),\pt\sgn(A)\}$ of the sign function and its derivative:
\begin{equation}
\label{eq:bA_error_est2}
 \epsilon_{\bA} = \sqrt{\left(\frac{\| \sgn(A)^2 \ket{\psi} - \ket{\psi} \|}{2 \ \| \ket{\psi} \|}\right)^2
 +
 \left(\frac{\| \{\sgn(A),\pt\sgn(A)\} \ket{\psi} \|}{2 \ \| \ket{\psi} \|}\right)^2 }.
\end{equation}
Note that the first summand in (\ref{eq:bA_error_est2}) is in general not equal to
$\epsilon_A^2$, because the optimal polynomials approximating the sign functions of $A$ and
$\bA$ are in general different. Since the square of the sign function is the
identity the anti-commutator $\{\sgn(A),\pt\sgn(A)\} = \pt(\sgn(A)^2)$ should vanish and its deviation from
zero can be used as an indicator of the precision with which the derivative $\pt \sgn(A)$
is calculated \cite{Buividovich:14:3}. We found numerically that $\epsilon_{\bA}$ gives a
better estimate for the true error of the derivative and moreover it is easier to compute
than the anti-commutator.

 During production runs it is not feasible to check the error for every source vector
$\ket{\psi}$.  In general the optimal subspace sizes will depend on the matrix and on the
source vector. If deflation techniques are used, however, the performance critical components
of the source vector are projected out and treated exactly. Then it is reasonable to
assume that for a given matrix the optimal parameter values will depend only weakly on the
source vector. This suggests that it is possible to find a set of optimal parameters that
will give the desired error for any vector. In order to find these optimal parameters we
perform a ``tuning run'' for every gauge configuration:

\begin{itemize}
\item Select a target precision $\epsilon_0$ (typically $\epsilon_0 = 10^{-8}$).
\item Select a trial vector $\ket{\phi}$. To compare different parameter sets the trial
  vector should be the same throughout the tuning run. Additionally it should be a good
  representation of the vectors that will be used in the production runs. Here we use
$\ket{\phi} = (1, \dots, 1)^\dagger$ to estimate the error $\epsilon_A$ of $\sgn(A)$ and the
sparse vector $\left(0, \ket{\phi}\right)$ to estimate the error $\epsilon_{\bA}$ of $\sgn(\bA)$.
\item Choose a set of trial parameters for the outer ($k^O$) and inner ($k^I$) Krylov subspace
  size $p=\{(k^O_1,k^I_1),\dots,(k^O_m,k^I_m)\}$. For every $p_i \in p$ compute
  $\epsilon_i := \epsilon(p_i)$ and the CPU time $t_i := t(p_i)$ the TSL approximation took.
\item Sort out all  $p_i$ for which $ \epsilon_i > \epsilon_o$
\item From the remaining parameter sets $p_j$  choose the one with the smallest $t_j$
\item Save the optimal parameters $k^O_0$ and $k^I_0$ for use in the production runs
\end{itemize}

Of course there is no guaranty that for some vector other than the trial vector
$\ket{\phi}$ the parameters $k^O_0$ and $k^I_0$ found in the tuning run will give an error
smaller than $\epsilon_0$.  One way to validate the results from the tuning run is to
perform a cross check:
\begin{itemize}
\item Use  $k^O_0$ and $k^I_0$ to compute the error for $l$ random vectors (in practice, $l = 20 \ldots 30$)
\item Check if the maximal error for all random vectors is smaller than $\epsilon_0$.
\end{itemize}
If the maximal error is found to be too large one can restart the tuning run with a
different trial vector to get better estimates for the optimal parameters. We found that
in most cases the error for the random vectors has the same order of magnitude as the
error for the trial vector. As a rule of thumb if one wants to achieve a precision
$\epsilon_0$ one should use the target precision $10^{-1}\epsilon_0$ for the trial run.

The TSL algorithm implicitly generates a polynomial approximation of the sign
function. The highest order of the approximation polynomial is given by the size of the
outer Krylov subspace. Another way to test the validity of the optimal parameter estimate
obtained in the trial run is to compare the results of the TSL approximation to some other
method. To this end we have set $\mu = 0$ and compared the relative error of the TSL
approximation with the error of the minmax polynomial approximation \cite{Giusti2002},
which works well when $H$ is a Hermitian operator.

\begin{figure}[h] \centering
  \includegraphics[width=0.495\linewidth]{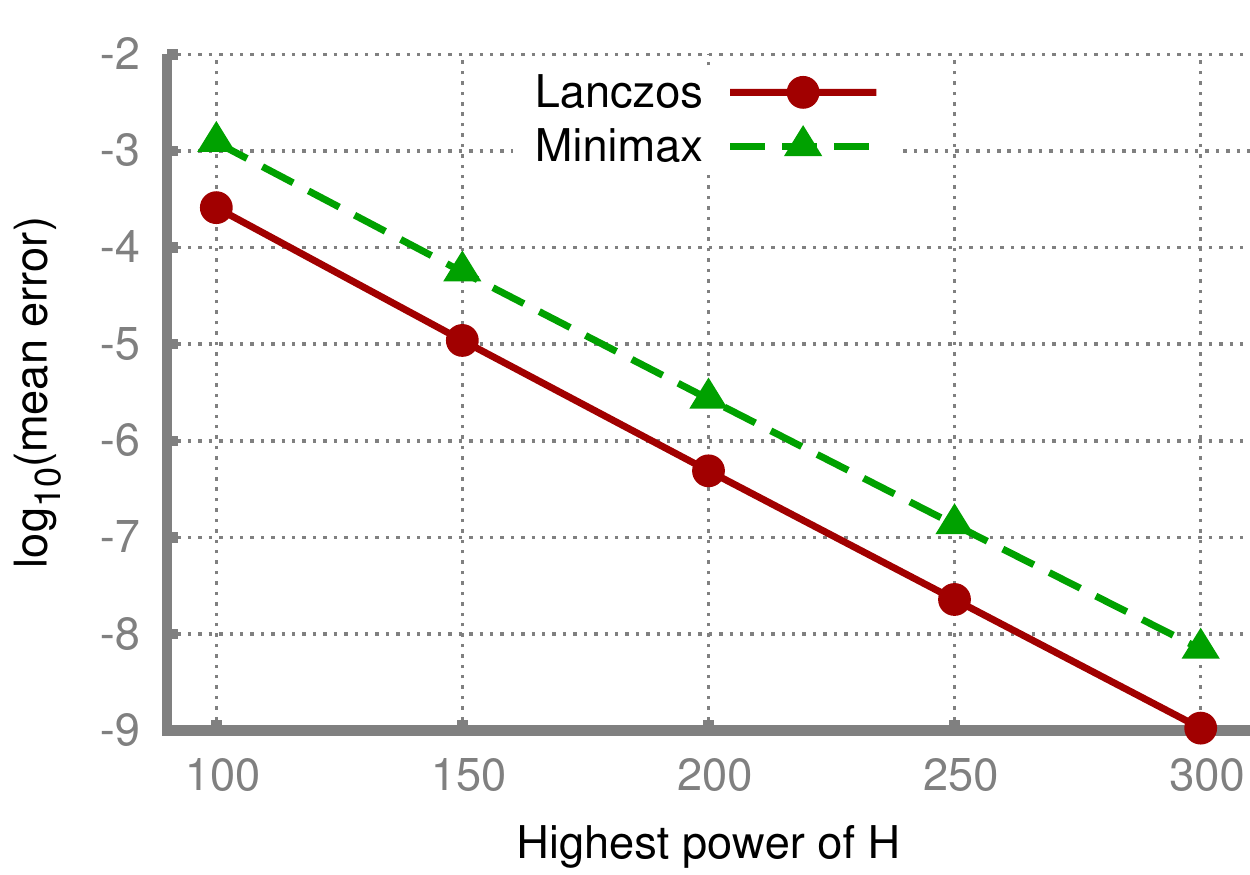}
  \includegraphics[width=0.495\linewidth]{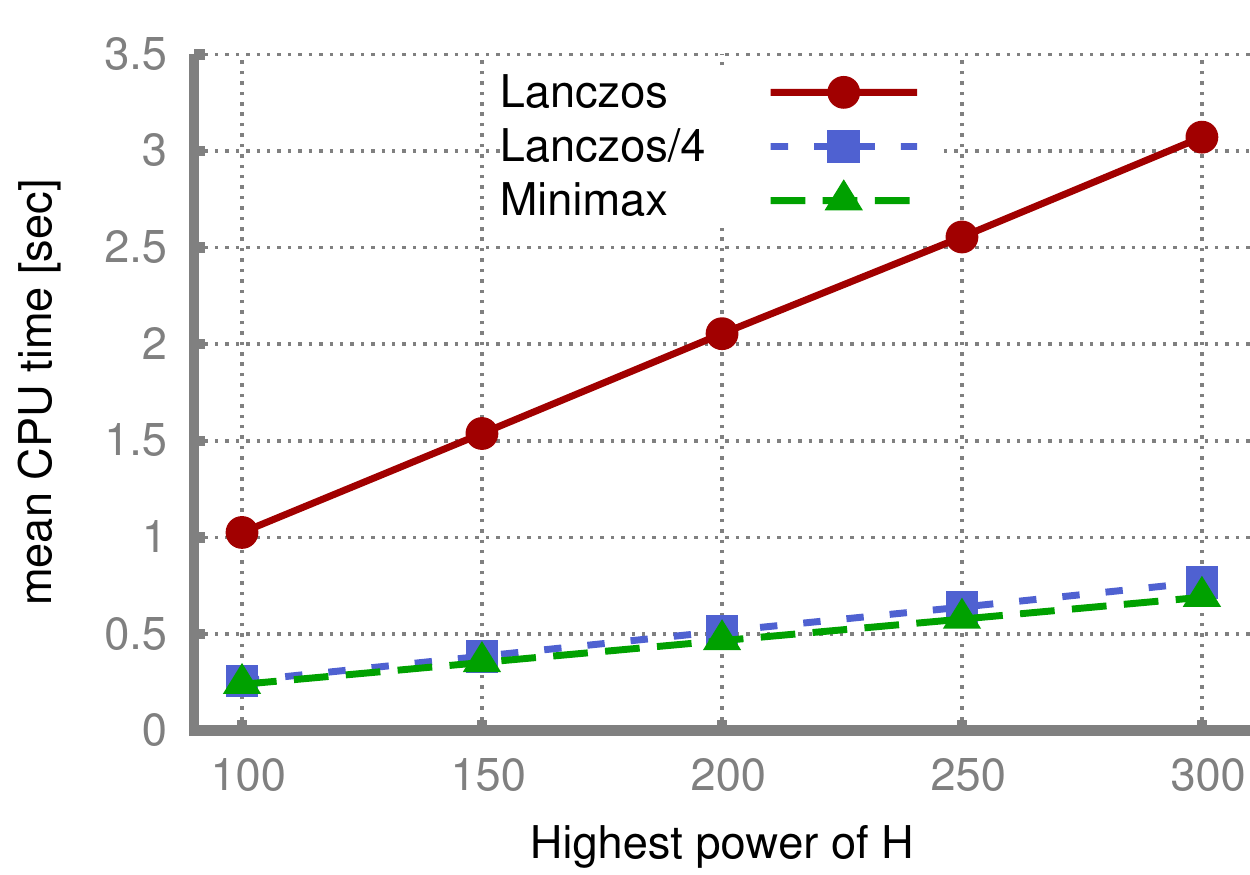}
  \caption{Comparison of the TSL approximation with the minmax polynomial method. On the
    left we show the mean error of both methods as a function of the highest power of the
    polynomial. The plot on the right shows the mean time required for the calculation of
    $\sgn(H) \ket{x}$ on an Intel\textsuperscript{\textregistered} Core\textsuperscript{TM}
    i5-3470 CPU, with multi-core OpenBLAS \cite{Xianyi} used for basic linear algebra. As 
    expected the TSL is slower than the minmax polynomial approach by a factor of four. }
  \label{fig:comp_minmax}
\end{figure}

 In Figure \ref{fig:comp_minmax} we compare the error of the TSL approximation for a given
outer Krylov subspace size with the minmax polynomial result. The error estimate
(\ref{eq:error_est}) for $\sgn(H)$ is computed for 10 random vectors on 20 different gauge
configurations of size $8\times 8^3$ (corresponding to a matrix of dimension $n=49152$)
with $\beta = 8.1$. For these parameters the spectrum of $H$ has only a small gap around
the line $\Re(z)=0$, which makes it more difficult to approximate the matrix sign function
with a polynomial of low order. To improve both the TSL and the minmax approximation we
deflated the $30$ eigenvalues with the smallest absolute value in our calculations.

The mean error is computed by averaging over random vectors and gauge configurations and
we plot the mean error as a function of the maximal power of $H$ in the approximating
polynomial (for the nested TSL algorithm, this is the size of the outer Krylov
subspace). The results are shown in Figure \ref{fig:comp_minmax}. We find that at fixed
polynomial degree the error for the TSL method is smaller than the error of the minmax
polynomial by almost an order of magnitude. This is to be expected, since the minmax
polynomial tries to minimise the maximal error over all vectors, while the TSL method
constructs a different and optimised polynomial for every source vectors. For large
matrices the main computational cost comes from matrix vector products and the overhead of
different algorithms for finding the optimal coefficients of the approximating polynomial
becomes negligible. The minmax polynomial method generates a polynomial in $H^2$, while
the TSL method constructs the Krylov subspaces for both $H$ and $H^{\dagger}$.  Moreover,
since storing all the Krylov vectors requires very large RAM memory and is hardly feasible
in practice, we have used a two-pass version of the TSL method. The first pass is used to
find the coefficients of the optimal approximating polynomial, and the second pass to
calculate this polynomial with known coefficients. Because of the twice larger number of
vectors used to construct the Krylov subspace and the need to calculate these vectors
twice, for a given order of the approximating polynomial the minmax approach is roughly
four times faster. The CPU time of the two algorithms is compared in
Figure~\ref{fig:comp_minmax}.
Our numerical test for $8\times 8^3$ configurations shows that the error of the TSL method
does not strongly depend on the source vector and that it is almost an order of magnitude
smaller that one would expect from a naive comparison with other approximation methods. For our
production runs we therefore use a single tuning run to find an estimate for the optimal
parameters. The parameter set found in this way is then used for all further calculations.

\subsection{TSL approximation for the derivatives of the sign function}
\label{subsec:tsl_approx_deriv}

 After tuning the TSL approximation for the sign function $\sgn(H)$, we are now ready to
apply it to the block matrix $\bA(H)$. Motivated by the practical calculations of
conserved vector currents on the lattice, we assume that the parameter $t$ is the Abelian
lattice gauge field $\Theta_{x,\mu}$ on the link which goes in the direction $\mu$ from
the lattice site $x$.

\begin{figure}[h]
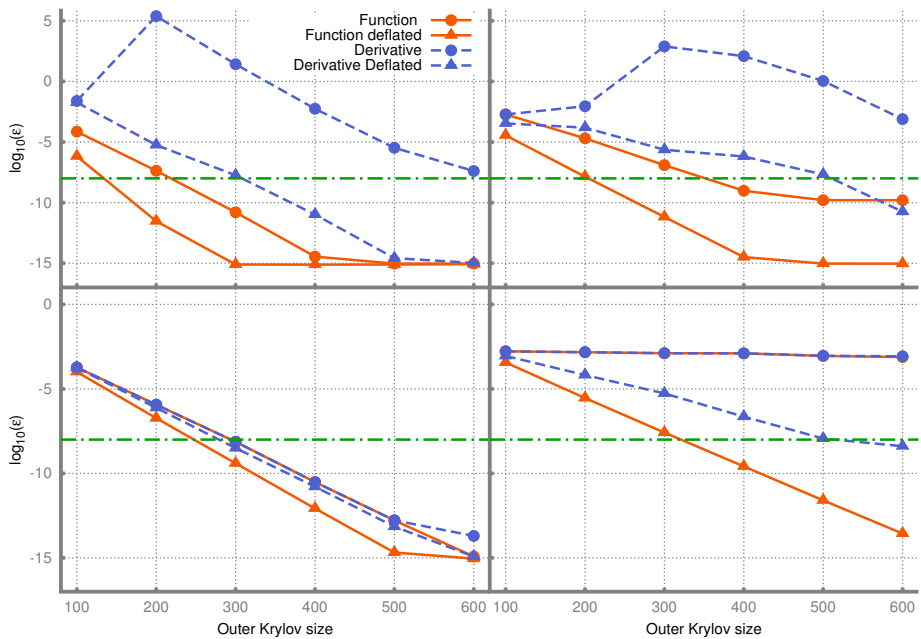
 \centering
   \includegraphics[width=0.99\linewidth]{{{fig2}}}
   \caption{Matrix size dependence of the optimal outer Krylov subspace size for zero
     chemical potential.  The green, dash-dotted line marks the desirable error of $10^{-8}$. (Un-)deflated
     results are marked by (circular) triangular symbols. The data points are connected with
     lines to guide the eye. A solid line stands for the sign function results while a broken
     line indicates the results for the derivative. Clockwise from top left the results are
     shown for $4\times4^3$, $6\times6^3$, $14\times14^3$ and $6\times18^3$ lattices. For
     $4\times4^3$, $6\times6^3$, $14\times14^3$ lattices we have used $\beta = 8.1$ ($a =
     0.125\fm$) and for $6\times18^3$ lattice we have used $\beta = 8.45$ ($a = 0.095\fm$). The
     inner Krylov subspace size is set to $100$ in all plots. For the deflation of the sign
     function we use the $40$ eigenvalues with the smallest magnitude. To deflate the
     derivative, we have used two eigenvalues for $4\times4^3$, $6\times6^3$ and $6\times18^3$
     lattices and six eigenvalues for $14\times14^3$.}
  \label{fig:krylov_comp}
\end{figure}

 In Figure \ref{fig:krylov_comp} we show typical results for the error of $\sgn(H)$ and
$\sgn(\bA(H))$ at $\mu=0$ as a function of the outer Krylov subspace size for
gauge configurations ranging in size from $4\times 4^3$ ($n=3072$) to $14\times14^3$
($n=460992$). The benefit of the deflation is clearly visible. In the computation of the
derivatives the deflation process has an additional positive side effect. To evaluate
derivatives we have to take a sparse input vector. Moreover the derivative matrix in
lattice QCD is very sparse, since the change of a single link influences only two lattice
sides. Therefore the vectors generated by the TSL algorithm have a sparse upper part and
the Krylov subspace does not efficiently approximate the full space. By projecting out the
vectors corresponding to the eigenvalues near zero the sparse pattern of the source vector
is destroyed and the Krylov subspace has a more general form which positively influences
the convergence rate of the method. Figure \ref{fig:krylov_comp} shows that the method
scales very well with the lattice volume. For the configuration with $V=6 \times 18^3$ and
$\beta = 8.45$ the temperature is already above the deconfinement transition temperature
for the Lüscher--Weisz action~\cite{Gattringer:02:1}, hence there is already a large
gap in the spectrum of $H$ and the deflation has only a minor effect on the efficiency of
the TSL method.

\begin{figure}[h]
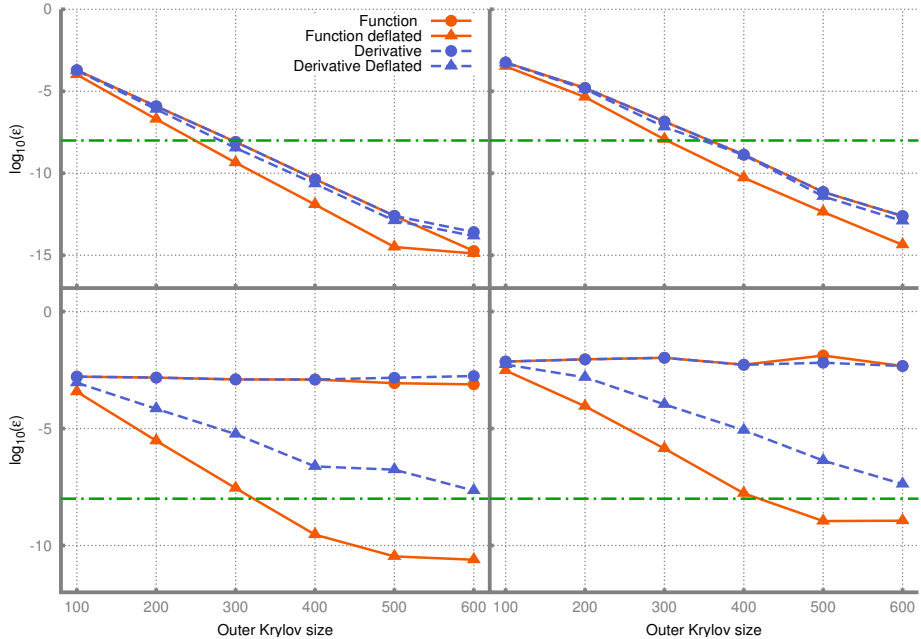

  \centering
   \includegraphics[width=0.99\linewidth]{{{fig3}}}
  \caption{Chemical potential dependence of the error as a function of the outer Krylov
    subspace size. Top: $V=6\times 18^3$ and $\mu=0.04$ on the left and $\mu=0.23$ on the
    right. Bottom: $V=14 \times 14^3$ and $\mu=0.05$ on the left and $\mu=0.3$ on the
    right. All other parameters are the same as for Figure \ref{fig:krylov_comp}.}
  \label{fig:krylov_comp_mu}
\end{figure}

An interesting question is how the optimal Krylov subspace size for a given error depends
on the chemical potential. As the chemical potential is increased the operator $H$
deviates more and more from a Hermitian matrix and one expects that a larger Krylov
subspace size is necessary to achieve a given accuracy. We indeed observe this behaviour
and Figure \ref{fig:krylov_comp_mu} shows the results for the error at different values of
$\mu$. From the plots we can estimate that at $\mu=0$ an outer Krylov subspace size of 500
is sufficient to obtain an error of $10^{-8}$ for the deflated derivative for the $V=14
\times 14^3$ configuration.
For $\mu=0.05$ one has to use a subspace size of roughly 600 for the same error. At
$\mu=0.3$ it seems that the error can be achieved with a subspace size of around
650. Going from zero chemical potential to a finite value of $\mu=0.05$
makes it necessary to increase the Krylov subspace size by about $20\%$. Once the chemical
potential is switched on, however, the increase in the Krylov subspace size is not that
dramatic. Between $\mu=0.05$ and the rather large value $\mu=0.30$ the increase in the
Krylov subspace size is roughly $10\%$.
For the  $V=6 \times 18^3$ configuration switching on a small chemical potential only has
a negligible effect on the error for a given Krylov subspace size. If $\mu=0.23$ one has
to take a Krylov subspace size of approximately 350 to get an error of $10^{-8}$. Compared
to the Krylov subspace size of 280 for $\mu=0$ this is an increase of about $25\%$.

\subsection{Divergence of $U\lr{1}$ vector current}
\label{subsec:divergence}

 As a further practice-oriented test of our method, we now consider the divergence of the
$U\lr{1}$ vector current
\begin{eqnarray}
\label{eq:divergence_def}
 \delta j_x = \sum\limits_{\mu} \lr{j_{x, \mu} - j_{x - \mu, \mu}}
\end{eqnarray}
at a randomly chosen lattice site $x$ for a fixed gauge field configuration which was
randomly selected from an ensemble of equilibrium gauge field configurations. For a fixed
gauge field configuration, the current $j_{x,\mu}$ flowing in the direction $\mu$ from the
lattice site $x$ is given by
\begin{equation}
 \label{eq:lat_vec_cur} j_{x,\mu} = \Tr \left(\Dov^{-1} \frac{\partial \Dov}{\partial
\Theta_{x,\mu}}\right) .
\end{equation}
Invariance of the Dirac operator under the gauge transformations ${\Theta_{x,\mu}
\rightarrow \Theta_{x, \mu} + \phi_x - \phi_{x + \mu}}$ demands that the divergence of the
vector current at a given lattice site vanishes.

 For small lattice sizes the trace in (\ref{eq:lat_vec_cur}) can be calculated exactly,
but for larger volumes this is no longer feasible. We use stochastic estimators with
$Z_2$-noise \cite{Dong:94:01} to compute an approximation of the trace for all currents in
(\ref{eq:divergence_def}). The error bars on our results are given by the standard
estimate for the error of the sample mean.

\begin{figure}[h]
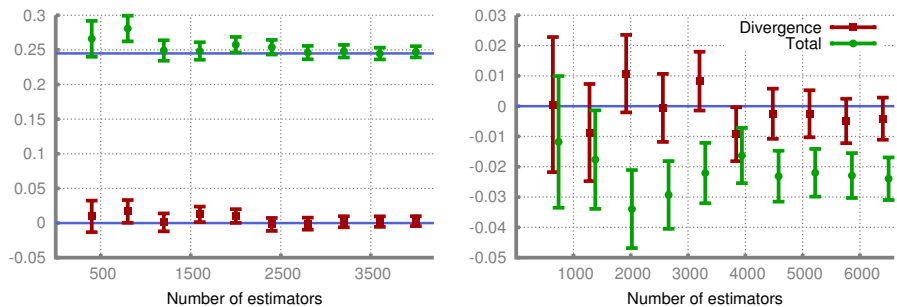

  \centering
  \includegraphics[width=0.49\linewidth]{{{fig4_a}}}
  \includegraphics[width=0.49\linewidth]{{{fig4_b}}}
  \caption{Results for divergence (red squares) of the vector current and the total
    current (green circles) at a lattice site for configurations with $\beta=8.1$ and
    $\mu=0.3$, plotted as functions of the number of stochastic estimators. On the left
    $V=3\times 4^3$. For this small configuration the exact values can be computed and are
    indicated by solid blue lines. On the right $V=6\times 6^3$. Here it is not feasible to
    compute the exact values. The results for the total current are shifted by 100 estimators
    to the right for better visibility and the blue line marks zero.}
  \label{3x4b8.1_divtot}
\end{figure}

 As a further check we additionally compute the ``total current'' at a lattice site, which
is just the sum over incoming and outgoing currents (replacing the minus sign in
(\ref{eq:divergence_def}) with a plus). This quantity has no physical meaning and can take any
value. If the total current is not zero an exact cancellation is necessary to achieve a
vanishing divergence. Finding numerically that the divergence vanishes even when the total
current does not can be seen as an additional cross check.  Figure~\ref{3x4b8.1_divtot}
shows the results for the divergence and the total current at finite $\mu$ for two
configurations of different size. For both configurations we find that the total current
has a finite value but the divergence vanishes as the number of stochastic estimators is
increased.

\begin{figure}[h]
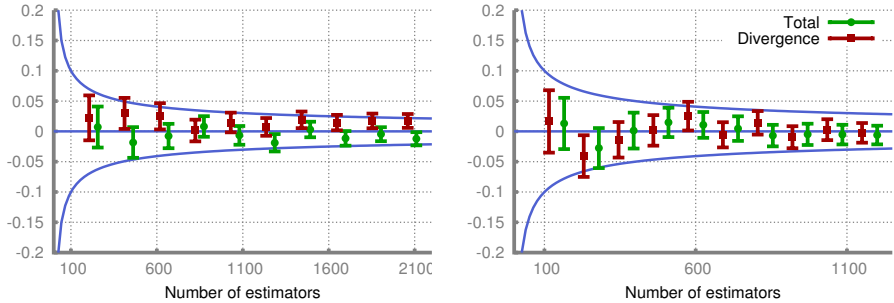

  \centering
  \includegraphics[width=0.49\linewidth]{{{fig5_a}}}
  \includegraphics[width=0.49\linewidth]{{{fig5_b}}}
  \caption{Results for divergence and the sum of the vector current at a lattice site for
large lattices, plotted as functions of the number $n_e$ of stochastic estimators. Left
$V=6\times18^3$ with $\beta = 8.45$, right $V=14\times14^3$ with $\beta = 8.1$. The
chemical potential is $\mu=0.0$ for both plots. The results for the total current are
shifted by 50 estimators to the right for better visibility. The straight blue line marks
zero and the curved blue lines depicting $\pm 1/ \sqrt{n_e}$ are drawn to visualise the
error dependence on the number of estimators.}
  \label{fig:large_divtot}
\end{figure}

Results for larger configurations are shown in Figure \ref{fig:large_divtot}. Computing a
large number of stochastic estimators is very expensive even for relatively small lattice
sizes. For the larger lattices we therefore studied the current conservation only for the
case $\mu=0.0$. We found that the divergence of the current as well as the total current
is very small for both configurations. To see a clear separation between the divergence
and the total current one would have to significantly increase the number of stochastic
estimators. The value of the divergence is in both cases consistent with zero and we
emphasise that the error bars clearly show the expected inverse square root dependence on
the number of estimators.

\section{Discussion and conclusions}
\label{sec:discussion}

In this work we have proposed and tested an efficient numerical method to compute derivatives of
matrix functions of general complex matrices. In particular we have shown that our method
in combination with the TSL approximation can be used to compute derivatives of the matrix
sign function with high precision. When one computes an approximation of a matrix
function it is often possible to improve the efficiency of the algorithm by deflating a
relatively small number of eigenvalues and treating them exactly. For this reason we have
also presented a generalised deflation method that does not depend on the
diagonalisability of the matrix.

In practical calculations it is important to know (an estimate for) the error of an
algorithm. We give a heuristic for tuning the parameters of our method to a given
error. To check how reliable the error estimate is we computed the error for a variety
of random source vectors on a number of medium sized ($8\times 8^3$) gauge
configurations. As a cross check we compared the errors of the TSL approximation with the
errors obtained by a minmax polynomial approximation. We found that our heuristic works
very well and gives a good error estimate. The error of the TSL method is consistently
lower than the error of the minmax polynomial by almost an order of magnitude. The TSL
method constructs two Krylov subspaces and we have to use a double-pass version of the
algorithm because of memory limitations, which makes the  TSL method about a
factor of four slower than the minmax polynomial for a given order of the interpolating
polynomial. We note that our findings suggest that a Lanczos type algorithm that is
optimised for Hermitian matrices could outperform the minmax polynomial approximation. The
Lanczos method for Hermitian matrices constructs only one Krylov subspace size, which
reduces the computational effort and the memory requirements by a factor two. Therefore a
single-pass version of the Lanczos algorithm could be feasible even for relatively large
Hermitian matrices. At a given precision, such an algorithm would be considerably faster
than the minmax approximation.

We have tested our method for the derivatives of the finite-density overlap Dirac operator
on different gauge configurations and for different values of the chemical potential
$\mu$. The higher the value of $\mu$ the more the operator $H$ differs from a Hermitian
matrix. We find that our method performs very well even for large values of $\mu$ up to
$470 \, \MeV$ in physical units and that the Krylov subspace size necessary to achieve a
given error does not strongly depend on $\mu$.  For all configurations we compared the
deflated and the undeflated version of the algorithm. With our choice of parameters in the
confining phase the operator $H$ in general only has a rather small gap in the spectrum
around the discontinuity line $\Re(z)=0$ and deflation greatly improves the performance of
the method. On the other hand we find numerically that in the deconfinement phase the gap
in the spectrum of $H$ is larger and there are no eigenvalues with real part close to
zero. In this case the deflation has only a minor effect.

In the deflated version of our method a large fraction of the overall CPU time is spent
for the inversion of the matrices $A_{\lambda_i} = \lambda_i - A$ in equation~(\ref{eq:c1i_2}). All the
inversions act on the same source vector and if it is possible to use a multi-shift
algorithm for the inversions the performance of the method could be significantly
improved. We tried to use the BiCGSTAB algorithm to compute $A_{\lambda_i}^{-1}$, since
multi-shift versions of this algorithm exist. It turned out, however, that this approach is
numerically unstable and in the end we computed  the inverse by applying the CG algorithm
to the matrix $A_{\lambda_i}A_{\lambda_i}^{\dagger}$. Thus finding a suitable multi-shift
inverter is one of the possible ways to speed up our algorithm.

Equivalents of  Theorem \ref{th:mathias} exist also for higher order derivatives of matrix functions
\cite{Mathias1996} and in principle it is straightforward to generalise our method for
higher derivatives. To compute the $k$-th derivative of a function of a $n$-dimensional
matrix one has to construct an upper-triangular block matrix of dimension $(k+1)n$. In
particular, second-order derivatives are required for the calculation of current--current
and charge--charge correlators, from which one can extract electric conductivity and charge
diffusion rate. However, in the case of higher derivative the expressions for the deflated
derivatives become extremely complicated. It seems thus that the only practical way to
calculate higher derivatives is to avoid deflation, which necessarily involves restriction
to small lattice sizes or to high temperatures.

An important application of our method and the main motivation for this work is the
computation of conserved currents for the finite-density overlap Dirac operator in lattice
QCD. Conserved currents and current-current correlators are important observables for the
study of anomalous transport effects such as the Chiral Magnetic \cite{Kharzeev:08:2} or
the Chiral Separation \cite{Son:06:2,Metlitski:05:1} effects. These dissipation-less
parity-odd transport effects, which originate from the chiral anomaly, have recently
become the focus of intensive studies in both high-energy and solid-state physics. While
the anomalous transport for free chiral fermions is well understood, there are still many
interesting open questions for strongly interacting fermions. With our method it is
possible to efficiently study anomalous transport for strongly interacting chiral fermions
on the lattice at finite quark chemical potential.

\section{Acknowledgements}
\label{sec:ack}
 We thank Andreas Frommer for pointing us to Theorem \ref{th:mathias}, and Jacques Bloch
and Simon Heybrock for useful discussions of the Lanczos approximation. This work was
supported by the S.~Kowalevskaja award from the Alexander von Humboldt Foundation.

\FloatBarrier
\appendix

\section{The Wilson--Dirac operator and its derivative}
\label{ap:WD_operator}
The Wilson--Dirac operator $\Dw(\mu)$ for a single quark flavour at finite quark chemical
potential $\mu$ and in the presence of a background $U(1)$ gauge field can be written as
\begin{equation}
  \label{eq:WD_op}
  \Dw(\mu) = \id - \kappa \sum\limits_{i=1}^3\left(G_i^+ + G_i^-\right) - \kappa
  \left(e^{\mu}G_4^+ + e^{-\mu}G_4^- \right),
\end{equation}
with
\begin{equation}
  \label{eq:WD_G}
  (G_\nu^\pm)_{x,y} := (1 \pm \gamma_{\nu})U_{\pm \nu}(x)e^{i\Theta_{\pm
      \nu}(x)}\delta_{x\pm \hat{\nu},y}.
\end{equation}
The $U_{\pm \nu}(x) \in SU(3)$ are (dynamical) lattice gauge fields and the factors
$e^{i\Theta_{\pm \nu}(x)} \in U(1)$ describe the (background) lattice gauge fields
corresponding to the external Abelian gauge field $\Theta_{\nu}(x)$. We set the lattice
spacing to one and define the hopping parameter $\kappa := 1/(2m_W + 8)$, where $m_W \in
(0,2)$ is the Wilson mass term. The matrices $\gamma_\nu$ are the Euclidean Dirac
matrices.

Computing the derivative of the Wilson--Dirac operator with respect to the Abelian gauge
field is straightforward and the result reads as

\begin{equation}
  \label{eq:dWD_op}
  \left(\frac{\partial \Dw}{\partial \Theta_{\nu}(z)} \right)_{x,y} = - i \kappa
  \left( \ (G_\nu^+)_{x,y} \  \delta_{x,z} - (G_\nu^-)_{x,y} \ \delta_{x-\hat{\nu},z} \ \right).
\end{equation}

\section{Derivatives of eigenvectors and eigenvalues}
\label{ap:derivatives}

In this Appendix we formally define the notion of the derivative of eigenvectors and
summarise some useful results.

Let $A(t)$ be a diagonalisable matrix which depends on a parameter $t$ and has
eigenvalues $\lambda_i$ and (left) eigenvectors ($\LL{i}$) $\RR{i}$.  For brevity we
assume that $A$ has no degenerate eigenvalues, i.e. $\lambda_i \neq \lambda_j$ if $i \neq
j$. For the Wilson--Dirac operator on real lattice QCD configurations, this is usually the
case, since configurations with degenerate eigenvalues form a set of measure zero in the
space of all gauge field configurations.

If $\RR{j}$ is an eigenvector of $A$ so is $\alpha \RR{j}$ for any $ \alpha \in
\mathbb{C}\setminus \{0\}$. The direction of an eigenvector is fixed, but its norm and
phase are not. In practice a common choice is to fix the eigenvectors by requiring that
the left and right eigenvectors are bi-orthonormal:
\begin{equation}
  \label{eq:eignorm}
  \braket{L_i}{R_j} = \delta_{ij}
\end{equation}
In general $\alpha$ can be any non-vanishing differentiable function $\alpha(t)$. The
freedom in choosing $\alpha$ leads to a freedom in the norm and the direction of the
derivative of the eigenvector, since
\begin{equation}
  \label{eq:eig_deriv_uncert}
  \pt  \left(\alpha \RR{j} \right) = \alpha {\dRR{j}} + (\pt \alpha) \RR{j}
\end{equation}
This means the derivative of an eigenvector is fixed only up to a multiplication with a
scalar and the addition of any vector from $\operatorname{span}(\RR{j})$. Therefore it is
necessary to specify which one of all this possible derivatives is used in a certain
calculation. Requiring the normalisation (\ref{eq:eignorm})  only leads to
the restriction $\Re\left(\braket{L_j}{\pt R_j}\right)=0$ and
does not fully fix the derivatives of the eigenvectors.  Throughout this paper we will
therefore employ the additional constraint
\begin{equation}
  \label{eq:eig_deriv_constr}
  \braket{L_j}{\pt R_j} = 0
\end{equation}
so that the derivative of an eigenvector is well defined. It is always possible to choose
the eigenvectors such that (\ref{eq:eig_deriv_constr}) is fulfilled. Note that condition
(\ref{eq:eignorm}) does not fix the norm of the vectors $\RR{j}$ and $\LL{j}$. Suppose we
found vectors that obey (\ref{eq:eignorm}). Then $\braket{L_j}{\pt R_j} = \xi $, where
$\xi$ is either zero or purely imaginary.  If we now define $\ket{\overline{R_{j}}}:=e^{-\xi
t}\RR{j}$ and $\bra{\overline{L_{j}}}:=\LL{j} e^{\xi t}$ we find that
$\braket{\overline{L_j}}{\overline{R_j}}  =  1 $  and $\braket{\overline{L_j}}{\overline{\pt R_j}}  =
0 $.

Using the definitions above we will now derive some useful relations. We start with the
eigenvalue equation
\begin{equation}
  \label{eq:eigval}
  A \RR{j} = \lambda_j \RR{j}.
\end{equation}
Taking the derivative on both sides yields
\begin{equation}
  \label{eq:eig_deriv}
  (\pt A)\RR{j} + A{\dRR{j}} = (\pt \lambda_j) \RR{j} + \lambda_j {\dRR{j}}.
\end{equation}
Multiplying from the left by $\LL{j}$ gives the following relation for the derivative of the eigenvalue:
\begin{equation}
  \label{eq:eigval_der}
  \LL{j}(\pt A)\RR{j} = {\pt \lambda}_j
\end{equation}
To derive a similar result for the derivative of the eigenvectors multiply
(\ref{eq:eig_deriv}) from the left by $\LL{i}$ for some $i \neq j$. With the
normalisation (\ref{eq:eignorm}) this gives
\begin{equation}
  \label{eq:eigenvec_deriv_1}
  \LL{i} (\pt A) \RR{j}  + \lambda_i  \braket{L_i}{\pt R_j}  = \lambda_j
  \braket{L_i}{\pt R_j}.
\end{equation}
Therefore the following equation holds for $i \neq j$:
\begin{equation}
  \label{eq:eigenvec_deriv_2}
   \braket{L_i}{\pt R_j} = \frac{\LL{i} (\pt A) \RR{j}}{\lambda_j - \lambda_i}
\end{equation}
Multiplying this equation by $\RR{i}$ from the left and summing over $i \neq j$ yields
\begin{equation}
  \label{eq:eigenvec_deriv_3}
  \dRR{j} = \sum\limits_{i \neq j}
\frac{\RR{i}\LL{i} (\pt A) \RR{j}}{\lambda_j - \lambda_i},
\end{equation}
where we used (\ref{eq:eig_deriv_constr}) and the identity
$\sum\limits_{i=1}^n \RR{i}\LL{i} = \mathbbm{1}$ to
make the replacement $\sum\limits_{i\neq j} \RR{i} \braket{L_i}{\pt R_j}
=\sum\limits_{i=1}^n \RR{i} \braket{L_i}{\pt R_j} =\dRR{j}$.
Similarly we obtain for the derivative of the left eigenvectors
\begin{equation}
  \label{eq:eigenvec_deriv_left}
  \dLL{j} = \sum\limits_{i \neq j} \frac{\LL{j} (\pt A) \RR{i}\LL{i}}{\lambda_j - \lambda_i}.
\end{equation}

\section{Properties of the block matrix $\bA$}
\label{ap:prop_mathias}
The convergence properties of matrix function approximation methods in general depend on
the spectrum of the matrix.  It is therefore important to know the spectrum of
$\bA$. Let $\bA$ be defined as in Theorem \ref{th:mathias} and let $\lambda_i$, $i = 1,
\cdots , n$ be the eigenvalues of $A$. Since $\bA$ is an upper block matrix $\det(\bA) =
\det(A)\det(A) = \det(A)^2$. From this it immediately follows that the eigenvalues of
$\bA$ are degenerate and identical to the $\lambda_i$.

A matrix is diagonalisable if and only if its minimal polynomial is a product of
\emph{distinct} linear factors. We will now discuss the outline of a proof of Theorem
\ref{th:diag}, which states that in general the matrix $\bA$ is not diagonalisable. In the
following we assume that $A$ has no degenerate eigenvalues to simplify the
argumentation. The generalisation to the case of degenerate eigenvalues is possible but a
bit more involved.  For every eigenvalue $\lambda_i$ of $A$ we define the two vectors
\begin{equation}
\label{eq:bA_evec}
  \ket{v_{i,1}} :=
  \begin{pmatrix}
    \RR{i} \\
     0
  \end{pmatrix} \quad \text{and} \quad
\ket{v_{i,2}} :=
  \begin{pmatrix}
    \dRR{i} \\
    \RR{i}
  \end{pmatrix}.
\end{equation}
It is easy to convince oneself that these vectors are linearly independent and that
$\ket{v_{i,1}}$ is an eigenvector of $\bA$ to the eigenvalue $\lambda_i$. For
$\ket{v_{i,2}}$ we have
\begin{eqnarray}
   \label{eq:bA_generalised_evec} \bA \ket{v_{i,2}} &=&
  \begin{pmatrix} A\dRR{i} + (\pt A)\RR{i} \\ A \RR{i}
  \end{pmatrix} =
  \begin{pmatrix} \pt (A \RR{i}) \\ A \RR{i}
  \end{pmatrix} \nonumber \\ \\ &=&
   \begin{pmatrix} \lambda_i \dRR{i} + (\pt \lambda_i)\RR{i} \\ \lambda_i \RR{i}
   \end{pmatrix} = \lambda_i \ket{v_{i,2}} + (\pt \lambda_i) \ket{v_{i,1}}. \nonumber
\end{eqnarray} The vector $\ket{v_{i,2}}$ is an eigenvector of $\bA$ only if $\pt
\lambda_i$ vanishes. If $\pt \lambda_i \neq 0$ we find
\begin{equation}
  \label{eq:bA_minus_lambda_kernel} (\bA - \lambda_i \id)^2 \ket{v_{i,2}} = (\pt
\lambda_i)(\bA - \lambda_i \id) \ket{v_{i,1}} = 0
\end{equation} and therefore $\ket{v_{i,2}}$ is a generalised eigenvector of rank two
corresponding to the eigenvalue  $\lambda_i$. From this and the fact that the algebraic
multiplicity of the eigenvalue $\lambda_i$ of $\bA$ is two it immediately follows that the
multiplicity of $\lambda_i$ in the minimal polynomial of $\bA$ is also two.
This proves the first part of Theorem~\ref{th:diag} .

To see that the second part of Theorem \ref{th:diag} is true, note that for every
eigenvalue of $A$ we have at least one Jordan block. Moreover the size of the largest
Jordan block belonging to an eigenvalue $\lambda_i$ is the multiplicity of the eigenvalue
in the minimal polynomial. Therefore if the eigenvalues are all pairwise distinct there
are at least $n$ Jordan blocks of size $2$ and since the dimension of $\bA$ is $2n$
this proves the Theorem.

\section{Derivation of the Jordan Decomposition}
\label{ap:jordan_decomp}
The aim of this appendix is to derive the analytic form of the Jordan decomposition of
$\bA$ in terms of the eigenvectors and eigenvalues of $A$. In practice one never
encounters a matrix $A$ with degenerate eigenvalues. Moreover if one or more of the
derivatives of the eigenvalues of $A$ vanishes the Jordan matrix of $\bA$ only becomes
simpler. Therefore in this paper we assume that every Jordan block of $\bA$ is of size
two.The generalisation to the case where some of the Jordan blocks have size one is
straightforward.

If $\mathcal{J}$ is the Jordan normal form of the matrix $\bA$ there exists an invertible matrix
$\mathcal{X}$ such that $ \mathcal{X}^{-1} \bA \mathcal{X} = \mathcal{J}$, i. e.
\begin{equation}
  \label{eq:jordan}
   \bA \mathcal{X} = \mathcal{X} \mathcal{J}
\end{equation}
The exact form of the Jordan matrix $\mathcal{J}$ follows form Theorem \ref{th:diag} and can be
exploited to compute the transformation matrix $\mathcal{X}$. In bra--ket notation the transformation
matrix reads as $\mathcal{X}:=(\ket{x_1}, \dots , \ket{x_{2n}})$.  Evaluating the right hand side of
equation
(\ref{eq:jordan}) yields
\begin{eqnarray}
  \label{eq:jordan_rhs1}
    \begin{pmatrix}
      x_{1,1} & \dots & x_{1,2n} \\
      &       &        \\
      \vdots &      & \vdots \\
      &      &       \\
      x_{2n,1} & \dots & x_{2n,2n}
    \end{pmatrix}
    \begin{pmatrix}
      \lambda_1 & 1          & 0      & \cdots    & 0      \\
      0         & \lambda_1  & \ddots & \ddots    & \vdots \\
      \vdots    & \ddots     & \ddots & \ddots    & 0      \\
      \vdots    &            & \ddots & \lambda_n & 1       \\
      0         & \cdots     & \cdots & 0         & \lambda_n
    \end{pmatrix} =  \nonumber \\
    \\
    \nonumber \begin{pmatrix}
      \lambda_1 x_{1,1} & x_{1,1} + \lambda_1 x_{1,2} & \dots &
      \lambda_n x_{1,(2n-1)} & x_{1,(2n-1)} + \lambda_n x_{1,2n} \\
      \vdots & \vdots  & & \vdots &\vdots \\
      \lambda_1 x_{2n,1} & x_{2n,1} + \lambda_1 x_{2n,2} & \dots &
      \lambda_n x_{2n,(2n-1)} & x_{2n,(2n-1)} + \lambda_n x_{2n,2n}
    \end{pmatrix}
\end{eqnarray}
Combining equations (\ref{eq:jordan}) and (\ref{eq:jordan_rhs1}) leads to $n$ coupled equations
\begin{eqnarray}
  \label{eq:eigenvalue}
  \bA \ket{x_{(2j-1)}} & = & \lambda_j \ket{x_{(2j-1)}} \\
  \label{eq:coupled}
  \bA \ket{x_{2j}} & = & \ket{x_{(2j-1)}} + \lambda_j \ket{x_{2j}},
\end{eqnarray}
where $j \in \{1,n\}$.
Equation (\ref{eq:eigenvalue}) is just the eigenvalue equation for the matrix $\bA$
and is easy to solve. Assume that $\RR{j}$ is an eigenvector of $A$ to the eigenvalue
$\lambda_j$, then it is straightforward to show that $(\RR{j},0)^{\text{T}}$ is an
eigenvector of $\bA$ to the same eigenvalue.

Using the solution of equation (\ref{eq:eigenvalue}) and defining
$\ket{x_{2j}}:= (\ket{x_{j,1}},\ket{x_{j,2}})^{\text{T}}$ equation (\ref{eq:coupled}) can be
written as
\begin{equation}
  \label{eq:sol}
  \begin{pmatrix}
    A & \pt A \\
    0 & A
  \end{pmatrix}
  \begin{pmatrix}
    \ket{x_{j,1}} \\
    \ket{x_{j,2}}
  \end{pmatrix} =
  \begin{pmatrix}
    \RR{j} \\
     0
  \end{pmatrix} + \lambda_j
  \begin{pmatrix}
    \ket{x_{j,1}} \\
    \ket{x_{j,2}}
  \end{pmatrix}
\end{equation}
which simplifies to

\begin{eqnarray}
  \nonumber (A - \lambda_j)\ket{x_{j,1}} + (\pt A) \ket{x_{j,2}} & = & \RR{j}   \qquad \text{\upperRomannumeral{1}} \\
  \label{eq:sol3} \\
   \nonumber (A-\lambda_j)\ket{x_{j,2}} & = & 0 \qquad \quad \text{\upperRomannumeral{2}}
\end{eqnarray}
Equation \upperRomannumeral{2} in the system (\ref{eq:sol3}) is again the eigenvalue
equation for $A$ and the solution is simply $\ket{x_{j,2}} = \kappa_i \RR{j}$, where
$\kappa_i$ is a finite complex number. Using this result equation (\upperRomannumeral{1})
becomes
\begin{equation}
  \label{eq:sol4}
   (A - \lambda_j)\ket{x_{j,1}} + (\pt A) \kappa_i \RR{j}  =  \RR{j}.
\end{equation}

Let $\{\LL{j}\}_{j=1,...,n}$ denote the set of left eigenvectors of $A$,
i.e. $\LL{j} A = \lambda_j \LL{j}$ and assume the normalisation
$\braket{L_j}{R_i}= \delta_{ij}$. Then $P:=\sum\limits_{i\neq
j}\frac{\RR{i}\LL{i}}{\lambda_i - \lambda_j}$ is well defined. Multiplying both
sides of equation (\ref{eq:sol4}) by $P$ yields
\begin{equation}
  \label{eq:sol5}
  \sum\limits_{i\neq j}\RR{i}\braket{L_i}{x_{j,1}} =\sum \limits_{i\neq j} \kappa_i
\frac{\RR{i}\LL{i}}{\lambda_i - \lambda_j} (\pt A) \RR{j}.
\end{equation}
The term $\sum\limits_{i\neq j}\RR{i}\LL{i}$ on the right hand side of equation
(\ref{eq:sol5}) is a projector to the space $ \mathbbm{1} - \RR{j}\LL{j}$ and using
equation (\ref{eq:eigenvec_deriv_3}) one finds that the right hand side is equal to
${\dRR{j}}$. It follows that the projection of $\ket{x_{j,1}}$ is equal to $\kappa_j
{\dRR{j}}$ and therefore
\begin{equation}
  \label{eq:sol6}
  \ket{x_{j,1}} = \kappa_j {\dRR{j}} + \gamma_j\RR{j},
\end{equation}
where $\gamma_j$ is a complex number.
Note that $\RR{j}$ is in the kernel of $(A-\lambda_j)$, which means we can add any
scalar multiple of $\RR{j}$ to the solution $\ket{x_{j,1}}$ of equation
(\upperRomannumeral{1}) and get another solution. Exploiting this freedom it is possible
to set $\gamma_j = 0$.  To compute the value of $\kappa_j$ simply multiply equation
(\upperRomannumeral{1}) by $\LL{j}$ from the left. This annihilates the first term on
the left hand side and we obtain
\begin{equation}
  \label{eq:sol7}
  \kappa_j \LL{j} (\pt A) \RR{j} = 1.
\end{equation}
Applying equation (\ref{eq:eigval_der}) gives $\kappa_j = 1/{{\pt \lambda}}_j$. Putting
everything together yields an analytic expression for the columns of the matrix $\mathcal{X}$:
\begin{eqnarray}
  \nonumber  \ket{x_{(2j-1)} } & =  & \quad \begin{pmatrix}
    \RR{j} \\
     0
  \end{pmatrix} \\
  \label{eq:final} & \\
  \nonumber  \ket{x_{2j}} & = &  \frac{1}{{\pt \lambda}_i} \begin{pmatrix}
    {\dRR{j}} \\
     \RR{j}
  \end{pmatrix}.
\end{eqnarray}
To compute the columns of $\mathcal{X}^{-1}$ we start with the equation
\begin{equation}
  \label{eq:inverse_1}
  \mathcal{X}^{-1} \bA = \mathcal{J} \mathcal{X}^{-1},
\end{equation}
which follows directly from equation (\ref{eq:jordan}). It turns out that it is
advantageous to consider the transpose of this equation
\begin{equation}
  \label{eq:inverse_2}
  \bA^T \mathcal{Y} = \mathcal{Y} \mathcal{J}^T,
\end{equation}
where  $\mathcal{Y}:=(\mathcal{X}^{-1})^T$  is introduced to simplify the notation. The right hand side
of equation (\ref{eq:inverse_2}) can be written as
\begin{eqnarray}
  \label{eq:jordan_rhs}
    \begin{pmatrix}
      y_{1,1} & \dots & y_{1,2n} \\
      &       &        \\
      \vdots &      & \vdots \\
      &      &       \\
      y_{2n,1} & \dots & y_{2n,2n}
    \end{pmatrix}
    \begin{pmatrix}
      \lambda_1 & 0          & \cdots & \cdots    & 0 \\
      1         & \lambda_1  & \ddots &           & \vdots \\
      0         & \ddots     & \ddots & \ddots    & \vdots \\
      \vdots    & \ddots     & \ddots & \lambda_n & 0 \\
      0         & \cdots     & 0      & 1 & \lambda_n
    \end{pmatrix} =  \nonumber \\
    \\
    \nonumber \begin{pmatrix}
      \lambda_1 y_{1,1} + y_{1,2} & \lambda_1 y_{1,2} & \dots &
      \lambda_n y_{1,(2n-1)} + y_{1,2n} & \lambda_n y_{1,2n} \\
      \vdots & \vdots  & & \vdots &\vdots \\
      \lambda_1 y_{2n,1} + y_{2n,2} & \lambda_1 y_{2n,2} & \dots &
      \lambda_n y_{2n,(2n-1)} + y_{2n,2n} & \lambda_n y_{2n,2n}
    \end{pmatrix}
\end{eqnarray}
The next steps are analogous to the derivation of the columns of $\mathcal{X}$ above. Let $\ket{y_i}$
be the $i-$th column of $\mathcal{Y}$, then (\ref{eq:jordan_rhs}) is equivalent to $n$ systems of
two equations:
\begin{eqnarray}
  \label{eq:inv_eigenvalue}
  \bA^T \ket{y_{2j}} & = & \lambda_j \ket{y_{2j}} \\
  \label{eq:inv_coupled}
  \bA^T \ket{y_{(2j-1)}} & = & \lambda_j \ket{y_{(2j-1)}} + \ket{y_{2j}},
\end{eqnarray}
Equation (\ref{eq:inv_eigenvalue}) is an eigenvalue equation and the solution is $
\ket{y_{2j}} = \eta_j \left( 0, \ket{R_j^T}\right)^T$, where $\ket{R_j^T}$ is the
eigenvector of $A^T$ to the eigenvalue $\lambda_j$ and $\eta_j$ is a scalar constant
that will be fixed later by requiring $ \mathcal{Y}^T\mathcal{X}=\mathbbm{1} $.

With the notation $\ket{y_{(2j-1)}}:=\left(\ket{y_{j,1}}, \ket{y_{j,2}}\right)^T$ we can
rewrite equation (\ref{eq:inv_coupled}) as a system of two equations:
\begin{eqnarray}
  \nonumber (A^T - \lambda_j)\ket{y_{j,1}} & = & 0 \qquad  \qquad \text{\upperRomannumeral{1}} \\
  \label{eq:inv_coupled1} \\
   \nonumber (\pt A^T)\ket{y_{j,1}} + (A^T-\lambda_j)\ket{y_{j,2}} & = & \eta_j \ket{R_j^T}
             \quad \text{\upperRomannumeral{2}}
\end{eqnarray}
Mimicking the steps used to solve the system (\ref{eq:sol3}) one finds the following
expressions for the columns of $\mathcal{Y}$:
\begin{eqnarray}
  \nonumber  \ket{y_{(2j-1)} } & =  & \frac{\eta_j}{{\pt \lambda}_j} \quad \begin{pmatrix}
    \ket{R_j^T} \\
     \ket{\pt R_j^T}
  \end{pmatrix} \\
  \label{eq:inv_final} & \\
  \nonumber  \ket{y_{2j}} & = &   \eta_j \begin{pmatrix}
    0 \\
     \ket{R_j^T}
  \end{pmatrix}.
\end{eqnarray}
 The rows of the inverse transformation matrix $\mathcal{X}^{-1} = \mathcal{Y}^T$ follow immediately from
equation (\ref{eq:inv_final}):
\begin{equation}
  \label{eq:inv_x}
  \mathcal{X}^{-1} =
  \begin{pmatrix}
    \frac{\eta_1}{{\pt \lambda}_1} \left(\LL{1}, {\dLL{1}} \right) \\
    \eta_1 \left(0, \LL{1}\right) \\
     \vdots  \\
     \frac{\eta_n}{{\pt \lambda}_n} \left(\LL{n}, {\dLL{n}} \right)  \\
     \eta_n \left(0, \LL{n}\right)
  \end{pmatrix},
\end{equation}
where we used the fact that $\ket{R_j^T}^T = \LL{j}$.

To find the values of the complex constants $\eta_j$ one has to evaluate the product
$\mathcal{X}^{-1}\mathcal{X}$.   In this computation one encounters only four different types of bra--ket products,
which are all shown in the following matrix product:
\begin{equation}
  \label{eq:inv_prod}
  \begin{pmatrix}
       \frac{\eta_i}{{\pt \lambda}_i} \left(\LL{i}, {\dLL{i}} \right) \\
       \eta_i \left(0, \LL{i}\right)
  \end{pmatrix}
  \begin{pmatrix}
    \begin{pmatrix}
      \RR{j} \\
       0
    \end{pmatrix} & ,
    \frac{1}{{\pt \lambda}_j}
    \begin{pmatrix}
       {\dRR{j}} \\
     \RR{j}
    \end{pmatrix}
  \end{pmatrix} =
  \begin{pmatrix}
    \frac{\eta_i}{{\pt \lambda}_i}\delta_{ij} & 0 \\
    0 & \frac{\eta_i}{{\pt \lambda}_j} \delta_{ij}
  \end{pmatrix}
\end{equation}
The entry below the diagonal is trivially zero and the super-diagonal entry vanishes
because $\braket{L_i}{\pt R_j}+\braket{\pt L_i}{R_j}=\pt
\left(\braket{L_i}{{R_j}}\right)=\pt\delta_{ij} = 0$. With the choice $\eta_i =
{\pt \lambda}_i$ the right hand side of equation (\ref{eq:inv_prod}) becomes the unit
matrix and the explicit form of the transformation matrices is given by equations
(\ref{eq:xinv_final}) and (\ref{eq:x_final}) in the main text.

\section{Efficient deflation of derivatives of the sign function }
\label{ap:efficient_defl}

 In this Appendix an efficient algorithm for the deflation of derivatives
of the sign function is developed.  Let $\RR{i}$ and $\LL{i}$ be the left and right
eigenvectors of $A$ to the eigenvalue $ \lambda_i $, respectively. Considering
equation~(\ref{eq:func_derivative_1}) it is tempting to exploit the sparsity of
$(0,\ket{x})^T$ to simplify the deflation calculations. However it
turns out that it is necessary to define the deflation for general
$\ket{\Phi} := (\ket{x_1},\ket{x_2})^T$. The reason is that the most convenient
way to estimate the error of the TSL approximation is via equation~(\ref{eq:error_est}),
i.e. we apply the TSL approximation twice and measure the deviation from unity. Therefore
we need a deflation method that works with general input vectors. In analogy to
equation~(\ref{eq:final_res}) we then get
\begin{eqnarray} \nonumber \sgn(\bA)
  \begin{pmatrix} \ket{x_1} \\ \ket{x_2}
  \end{pmatrix} & = & \underbrace{\sum\limits_{i=1}^{k} s_i\left\{
    \begin{pmatrix} \RR{i} \\ 0
    \end{pmatrix} \left[\braket{L_i}{x_1}+\braket{\pt L_i}{x_2}\right] +
\begin{pmatrix} \dRR{i} \\ \RR{i}
\end{pmatrix} \braket{L_i}{x_2} \right\}}_{\text{\upperRomannumeral{1}}} \\
 \label{eq:array_app_defl_1} & & + \underbrace{\sgn(\bA)\bar{\mathcal{P}}_{2k}\begin{pmatrix}
\ket{x_1} \\ \ket{x_2}
  \end{pmatrix}}_{\text{\upperRomannumeral{2}}},
\end{eqnarray} where we used the fact that $\pt \sgn(\lambda_i)=0$ since the sign function is piece-wise
constant and introduced the notation $s_i:=\sgn(\lambda_i)$.
 Let us now investigate part~\upperRomannumeral{1} of (\ref{eq:array_app_defl_1}).
\begin{eqnarray}
  \label{eq:app_defl_2} \text{\upperRomannumeral{1}} & = & \sum\limits_{i=1}^{k}
s_i\left\{
    \begin{pmatrix} \RR{i} \\ 0
    \end{pmatrix} \underbrace{\left[\braket{L_i}{x_1}+\braket{\pt
L_i}{x_2}\right]}_{c_{1i}} +
\begin{pmatrix} \dRR{i} \\ \RR{i}
\end{pmatrix} \underbrace{\braket{L_i}{x_2}}_{c_{2i}} \right\}
\end{eqnarray} In a practical calculation we are interested in finding an efficient way to
compute the coefficients $c_{1i}$ and $c_{2i}$. Note that these coefficients are
proportional to the scalar products of $\ket{\Phi}$ with the odd and even rows of
the matrix $\mathcal{X}^{-1}$ respectively. The same coefficients appear in the projection
$\bar{\mathcal{P}}_{2k}\ket{\Phi}$, which is needed to compute part
\upperRomannumeral{2}. Apart from $c_{1i}$ and $c_{2i}$ the only non-trivial part of the
deflation is the computation of $\dRR{i}$.  As we mentioned earlier, the left and
right eigenvectors $\LL{i}$ and $\RR{i}$ can be computed with {\tt{ARPACK}}
routines.

The coefficients $c_{2i}$ are simply scalar products. As we will see later on they appear
in several parts of the deflation and therefore it pays off pre-compute and save them.

The first part of the coefficients $c_{1i}$ is again a scalar product. The computation of
the second part is more involved. First we use equation \ref{eq:eigenvec_deriv_left} to
get rid of the derivative of the eigenvector:
\begin{equation}
  \label{eq:c1i_1} \braket{\pt L_i}{x_1} = \sum\limits_{j \neq i } \frac{\LL{i}(\pt A)
\RR{j}\braket{L_j}{x_2}}{\lambda_i-\lambda_j}
\end{equation} Computing all the eigenvectors of $A$ is in general way too expensive and
only the first $k$ eigenvectors are known explicitly.  The trick now is to use the
identity $\sum\limits_{j=k+1}^n \frac{\RR{j}\braket{L_j}{x_2}}{\lambda_i - \lambda_j}
\equiv \left(\lambda_i - A\right)^{-1}P_k \ket{x_2}$, where $P_k :=
\sum\limits_{i={k+1}}^n \RR{i}\LL{i}$. Note that the inverse of $\left(\lambda_i -
A\right)$ is well defined for $P_k \ket{x_2}$ and we can write
\begin{equation}
  \label{eq:c1i_2} \braket{\pt L_i}{x_1} = \sum\limits_{\substack{j=1\\j \neq i}
}^{k}\frac{\LL{i}(\pt A) \RR{j}c_{2j}}{\lambda_i-\lambda_j} + \LL{i} (\pt A)
\left(\lambda_i - A\right)^{-1}P_k\ket{x_2}
\end{equation} In practical applications the matrix $A_{\lambda_i} := \left( \lambda_i - A
\right)$ is sparse and its inverse can be computed very efficiently with iterative
methods. Since the vector $P_k\ket{x_2}$ in equation (\ref{eq:c1i_2}) is the same for all
$\lambda_i $ with $i \in \{1, \dots, k\}$ it is in principle possible to use a multi-shift
inversion algorithm to compute the inversions. In practice, however, we found
that the numerical inversion with the (multi-shift) BiCGSTAB algorithm was unstable. To
avoid stability issues we used the CG algorithm to find the inverse of the Hermitian
matrix $A_{\lambda_i}A_{\lambda_i}^\dagger$, from which it is straightforward to compute
the inverse of $A_{\lambda_i}$.

The vector $\RR{i}$ lies in the kernel of the matrix $A_{\lambda_i}$. For this reason and
because of numerical errors the vector $A_{\lambda_i}^{-1}P_k\ket{x_2}$ can have
non-zero components in $\RR{i}$ direction. Remember that we normalised the derivative
of the eigenvectors such that $\braket{\pt L_i}{R_i} = 0$. In order to enforce this
normalisation in a numerical calculation we have to project out the spurious $\RR{i}$
component. To this end we define the projection operator $Q_i$ in the following way
\begin{equation}
  \label{eq:Q} Q_i\ket{\psi} := \ket{\psi} - \RR{i}\braket{L_i}{\psi}.
\end{equation} With this operator we can now write down the final equation for $
\braket{\pt L_i}{x_1}$ that can be used in numerical calculations
\begin{equation}
  \label{eq:c1i_3} \braket{\pt L_i}{x_1} = \LL{i} (\pt A) \left(
\sum\limits_{\substack{j=1\\j \neq i} }^{k}\frac{ \RR{j}c_{2j}}{\lambda_i-\lambda_j} +
Q_i\left(\lambda_i - A\right)^{-1}P_k\ket{x_2}\right).
\end{equation}
Analogously one can use equation (\ref{eq:eigenvec_deriv_3}) to
derive the following formula for the derivative of the right eigenvectors
\begin{equation}
  \label{eq:dRi} \dRR{i} = \sum\limits_{\substack{j=1\\j \neq i}
}^{k}\frac{\RR{j}\bra{L_j}(\pt A) \RR{i}}{\lambda_i-\lambda_j} + Q_i \left( \lambda_i
- A \right)^{-1} P_k (\pt A) \RR{i}.
\end{equation} We now have all the parts needed for an efficient computation of the
deflation. The whole computation is summarised in the following code listing:

\begin{center}
  \begin{algorithmic}[1]
    \Function{DeflatedSignDerivative}{} \State  // Compute
    $\ket{out} = (\ket{out_1},\ket{out_2})^T =
    \sgn(\bA)(\ket{x_1},\ket{x_2})^T$ \State // with
    deflation \State
    \For{$i \gets 1$
      \textbf{ to } $k$}  // Compute $\dRR{i}$
    \State $\dRR{i} = \ket{0}$
    \For{$j \gets 1$
      \textbf{ to } $k$} \If{$i \neq j$}
    \State
    $\dRR{i} = \dRR{i} + \frac{\RR{j}\bra{L_j}(\pt A)
      \RR{i}}{\lambda_i - \lambda_j}$
 \EndIf
    \EndFor
    \State
    $\dRR{i} = \dRR{i} + Q_i\left(\lambda_i-A\right)^{-1}P_k(\pt A)
    \RR{i}$
    \EndFor
    \State
    \For{$i \gets 1$
      \textbf{ to } $k$}  // Compute $c_{2i}$
    \State $c2[i] = \braket{L_i}{x_2}$
    \EndFor
    \State \State // Compute $c_{1i}$,
    the exact part and the projection of $\ket{x}$
    \State $\ket{x_k} = P_k \ket{x_2}$
    \State $\ket{out_1} = \ket{0} $
    \State $\ket{out_2} = \ket{0} $
    \For{$i \gets 1$
      \textbf{ to } $k$}
    \State$ \ket{v} = Q_i\left(\lambda_i-A\right)^{-1} \ket{x_k}$
    \For{$j \gets 1$
      \textbf{ to } $k$} \If{$i \neq j$}
    \State $\ket{v} = \ket{v} + \frac{c2[j]\RR{j}}{\lambda_i-\lambda_j}$
    \EndIf
    \EndFor
    \State
    $c1[i] = \LL{i}\left((\pt A) \ket{v} + \ket{x_1}\right)$
    \State  // Exact part of output \State
    $s_i = \sgn(\lambda_i)$
    \State
    $\ket{out_1} = \ket{out_1} +
    s_i\left(c1[i]\RR{i}+c2[i]\dRR{i}\right)$ \State
    $\ket{out_2} = \ket{out_2} + s_i\left(c2[i]\RR{i}\right) $
    \State  // Projection of input vector \State
    $\ket{x_1} = \ket{x_1} -
    \left(c1[i]\RR{i}+c2[i]\dRR{i}\right)$ \State
    $\ket{x_2} = \ket{x_2} - \left(c2[i]\RR{i}\right) $
    \EndFor
    \State // Exact part plus TSL approximation of projected part
    \State
    $\ket{out} = (\ket{out_1},\ket{out_2})^T +
    \TSL(\bA,(\ket{x_1},\ket{x_2})^T)$
    \State \Return $\ket{out}$
    \EndFunction
  \end{algorithmic}
\end{center}

\newpage

\end{document}